\documentstyle[aps,multicol,prl,epsfig]{revtex}
\newcommand{\Zeff}{Z_{\hbox{\scriptsize eff}}}
\newcommand{\Zsat}{Z_{\hbox{\scriptsize sat}}}

\newcommand{\Phipb}{\Phi_{_{\hbox{\scriptsize PB}}}}
\newcommand{\Philpb}{\Phi_{_{\hbox{\scriptsize LPB}}}}

\begin{document}
\title{A simple approach for charge renormalization of highly charged 
macro-ions}
\author{Emmanuel Trizac$^{1}$, Lyd\'eric Bocquet$^{2}$ and Miguel Aubouy$^{3}$}
\address{$^{1}$ Laboratoire de Physique Th\'eorique, 
UMR CNRS 8627, B{\^a}timent 210, Universit{\'e}  Paris-Sud,
91405 Orsay Cedex, France, \\
$^{2}$ Laboratoire de Physique de l'E.N.S. de Lyon,
UMR  CNRS 5672, 46 All\'ee d'Italie, 69364 Lyon Cedex, France \\
$^{3}$ S.I.3M., D.R.F.M.C., CEA-DSM Grenoble, 
17 rue des Martyrs, 38054 Grenoble Cedex 9, France}

\date{\today}
\maketitle

\begin{abstract}
We revisit the popular notion of effective or renormalized charge, which is
a concept of central importance in the field of highly charged colloidal or
polyelectrolyte solutions. Working at the level of a linear Debye-H\"uckel
like theory only, we propose a simple, efficient and versatile method
to predict the saturated amount of charge renormalization, which is however
a non-linear effect arising at strong electrostatic coupling. The results
are successfully tested against the numerical solutions of
Poisson-Boltzmann theory for polyions of various shapes (planar, cylindrical 
and spherical), both in the
infinite dilution limit or in confined geometry, with or without added
electrolyte. Our approach, accurate for monovalent
micro-ions in solvents like water, is finally confronted against experimental
results, namely the crystallization of charged colloidal
suspensions and the osmotic coefficient of B-DNA solutions.
\end{abstract}


\begin{multicols}{2}
\narrowtext 
Our present understanding of charged macro-ions suspensions 
is essentially based
on the DLVO theory, named after Derjaguin, Landau, Verwey and Overbeek \cite
{Verwey}. This approach relies on a Poisson-Boltzmann 
(PB) mean-field  description of the micro-ions clouds. 
An important prediction of the theory is the effective 
interaction pair potential between two macro-ions (e.g. colloids) in the solvent 
which, within a linearization approximation,   
takes the well-known Yukawa or Debye-H\"{u}ckel (DH)
form: $v(r)\sim Z^{2}\exp (-\kappa r)/r$, where $Z$ is the charge of the
object and $\kappa $ denotes the inverse Debye screening length. 
However, this approach becomes inadequate to describe highly charged 
objects for which
the electrostatic energy of a micro-ion near the macro-ion 
surface largely exceeds $k_{B}T$, 
the thermal energy, and the linearization of PB equations is a priori
not justified. In this case however, the electrostatic potential in
exact \cite{Kjellander} or mean-field 
\cite{Manning,Belloni,Alex} theories still
takes a Debye-H\"{u}ckel form far from the charged bodies, provided
that the source of the potential is renormalized ($Z\to Z_{%
\hbox{\scriptsize
eff}}$). The essential idea is that the micro-ions which suffer a high
electrostatic coupling with the macro-ion accumulate in its immediate vicinity
so that the decorated object (macro-ion {\it plus} captive counter-ions) may
be considered as a single entity which carries an effective charge $Z_{%
\hbox{\scriptsize eff}}$, much lower (in absolute value) than the structural
one. Within PB theory, $Z$ and $Z_{\hbox{\scriptsize eff}}$ coincide for
low values of the structural charge, but $Z_{\hbox{\scriptsize eff}}$
eventually reaches a saturation value $Z_{\hbox{\scriptsize sat}}$ 
independent of $Z$ when the
bare charge increases \cite{Alex,JPH}.

Of course, the difficulty remains to predict $Z_{\hbox{\scriptsize sat}}$
for a given suspension of macro-ions \cite{Belloni,Alex,JPH,Diehl}. 
In the absence
of any general analytical framework for the computation of the effective
charge, this quantity is often considered as an adjustable parameter to fit
experimental data \cite{Gast,Robbins2}. In this Letter we show that a simple
physical argument, at the level of the DH linearized description only,
yields explicit (and in some favorable cases analytical) expressions for
the effective charges at saturation, which compare well with both numerical 
solutions of non-linear PB theory and available experimental data.

For simplicity, we start by considering a unique highly (positively) 
charged sphere immersed
in a symmetric 1:1 electrolyte of bulk ionic strength $I_0=\kappa^2/(8 \pi 
\ell_B)$, where $\ell _{B}=e^{2}/(4\pi \epsilon
k_{B}T)$ is the Bjerrum length ($\epsilon$ being the dielectric constant
of the solvent): $\ell _{B}=7\,$\AA\ for
water at room temperature.
Within PB theory, the dimensionless electrostatic potential 
$\Phi=e V/k_BT$ obeys the relation
\begin{equation}
\nabla^2 \Phi \,=\, \kappa^2 \,{\rm sinh} \Phi.
\label{PB}
\end{equation}
Far from macro-ion (where it is understood that $\Phi$ vanishes),
the solution  $\Phipb$ of Eq. (\ref{PB}) 
also obeys the linearized Poisson-Boltzmann (LPB) equation
$\nabla^2 \Phi = \kappa^2 \Phi$, and therefore takes the Yukawa form
$\Philpb = \Zeff \ell_B \exp[\kappa(a-r)]/[r(1+\kappa a)]$, with $a$ the radius
of the sphere.
$\Zeff$ (in $e$ units) is consequently defined here 
without ambiguity from the far
field behaviour of $\Phipb$ (see \cite{Belloni,Deserno,Lukatsky} 
for alternative definitions
of effective charges).
Accordingly, a ``non linear'' region may be defined ($r\in [a,r^*]$), corresponding to
 $\Phipb$ larger than unity,
where by definition of the cutoff $r^*$, $\Philpb(r^*)$ is of order 1.
In the limit of large $\kappa a$, this ``non-linear'' region
is however confined to the immediate vicinity of the macro-ion: 
$r^* \simeq a$.
We consequently have the effective boundary condition
$\Philpb(a) \simeq \Phipb(r^*) = {\cal C}$,
where ${\cal C}$ is a constant of order 1, which yields immediately
$\Zeff= {\cal C} a (1+ \kappa a)/\ell_B $. This argument assumes
that the bare charge $Z$ is high enough to have $\Phipb$ larger than unity
close to the macro-ion, and therefore provides the saturation value of
$\Zeff$, denoted hereafter as $\Zsat$ \cite{Rqe}.
We therefore easily obtain the non trivial dependence of
this quantity upon physico-chemical parameters.

This picture of a decorated macro-ion --where the ``bound'' counterions 
renormalizing the charge appear to have an electrostatic energy $eV_0$
balancing the thermal energy $k_BT$-- may be rationalized as follows.
In the limit of large $\kappa a$, we perform an asymptotic matching
of the non-linear PB planar solution (see \cite{Verwey}) 
to the linear solution $\Philpb$ in 
curved geometry. We obtain for high bare charges the same value 
of the contact potential $\Philpb(a) = 4$ (of order 1 as expected) so that 
$\Zsat= 4 a (1+ \kappa a)/\ell_B $ \cite{BTA}. 
Such a procedure provides by construction 
the correct large $\kappa a$ (low curvature) behaviour of $\Zsat$,
but we will show below that it remains fairly accurate down to $\kappa a$ of
order 1.

Generalizing this approach, we consequently obtain the leading curvature
saturated effective charge from the following analysis.
For an isolated macro-ion of arbitrary shape in an electrolyte: 
{\it a)} find the
 electrostatic potential, $\Phi _{_{\hbox{\scriptsize LPB}}}$,
solution of the{\em \ linearized} PB equations, supplemented by a  
{\em fixed potential boundary condition}: $\Phi
_{_{\hbox{\scriptsize LPB}}}(\hbox{surface})={\cal C}$, where ${\cal C}=4$ 
at leading order in curvature; {\it b)} 
deduce $Z_{\hbox{\scriptsize sat}}$ from  Gauss theorem at the surface of the
object. 
% Once the effective charge at saturation is known, a crude estimation
% for $Z_{\hbox{\scriptsize eff}}$ consists in setting $Z_{%
% \hbox{\scriptsize eff}}=Z$ for $Z<Z_{\hbox{\scriptsize sat}}$ and 
% $Z_{\hbox{\scriptsize eff}}=Z_{\hbox{\scriptsize sat}}$ for 
% $Z>Z_{\hbox{\scriptsize sat}}$.
In the case of an infinite cylinder (radius $a$, bare lineic charge $\lambda $), 
we obtain 
$\lambda _{\hbox{\scriptsize sat}}=
2(\kappa a/\ell _{B})K_{1}(\kappa a)/K_{0}(\kappa a)$, where
$K_{0}$ and $K_{1}$ are the modified
Bessel functions of orders 0 and 1. 
%%%%%%%%Fig 1%%%%%%%%%%   
\begin{figure}[tbh]
\epsfig{file=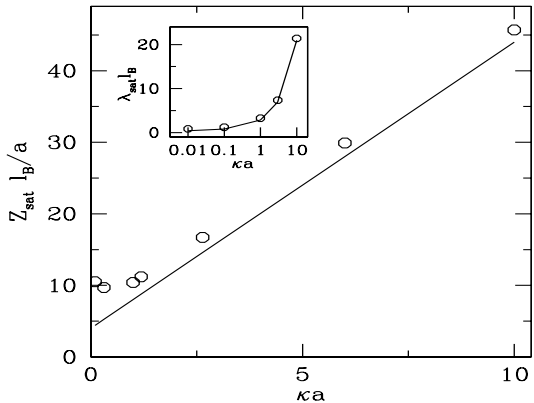,width=7.5cm,height=6.0cm}
\caption{Effective charge at saturation of an isolated spherical 
macro-ion (radius $a$) as a function of $%
\kappa a$. The continuous line is the analytical expression given in the text, 
while the dots are the results extracted from the far field
behaviour of the non-linear PB potential. In the inset, the same results for the
cylinder geometry are shown on a log-linear scale.}
\label{Charge_seule}
\end{figure}
%%%%%%%%%%%%%%%%%%%%%%%  

In order to
test the validity of our results, we have numerically solved the non-linear
PB equation (\ref{PB}) for high $Z$ values corresponding to the saturation regime, 
and computed the effective charge from the
electrostatic potential at large distances (i.e. the value required to match 
$\Phi _{_{\hbox{\scriptsize LPB}}}$ to the far field $\Phi _{_{%
\hbox{\scriptsize PB}}}$ obtained numerically). Figure \ref{Charge_seule}
compares the resulting PB effective charge to our
expressions, for spherical and cylindrical
macro-ions. The agreement becomes excellent at large $\kappa a$ as it should,
and in the case of cylinders, even holds down to
very small $\kappa a$ (0.01), a point which is not 
{\it a priori} expected. Finally, in the planar geometry
our approach provides by construction 
the correct effective charge (compared to PB).
% in the limit of vanishing salt $\kappa a\to 0$, the
% exact PB result for $\lambda _{\hbox{\scriptsize sat}}\ell _{B}$ is indeed
% finite ($2/\pi $ from the analytical work reported in \cite{Tracy}), while
% our estimate vanishes, although very slowly as $-1/\log (\kappa a)$.

%%%%%%%%Fig 2%%%%%%%%%%   
\begin{figure}[htb]
\epsfig{file=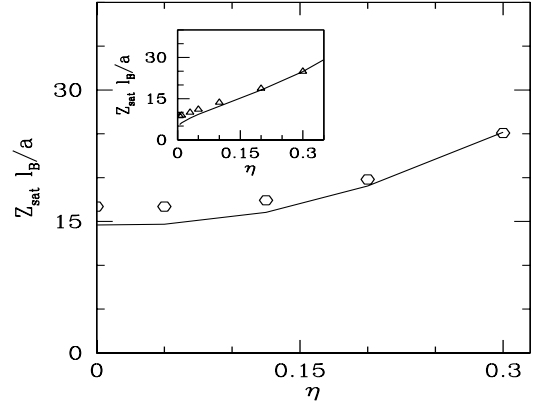,width=7.5cm,height=6.0cm}
\caption{ Effective saturated charge of spherical macro-ions (radius $a$) 
as a function of volume
fraction $\eta$ for $\kappa a=2.6$. The continuous line shows the 
effective charge computed
using the prescription, while the dots are the results of the non-linear PB
theory, following Ref. {\protect\cite{Alex}}. Inset: no-salt situation.}
\label{Effective_Charge}
\end{figure}
%%%%%%%%%%%%%%%%%%%%%%%

For spherical colloids, expressions reminiscent of that reported
above can be found in the literature \cite{Belloni,Crocker,Behrens}. 
It seems however that the generality of the underlying method 
has been overlooked. In particular, 
our procedure may be extended to the finite concentration case, using
the concept of Wigner Seitz (WS) cell \cite{Alex}: the influence of
the other macro-ions is accounted for by confining the macro-ion into a cell,
with global electroneutrality. The size of the cell, $R_{WS}$, is computed
from the density of macro-ions, while its geometry is chosen as to mimic the
spatial structure of the macro-ions in the solution. Suppose that the system
is in equilibrium with a reservoir of salt defined in terms of its
Debye length $\kappa ^{-1}$. We first linearize Eq. (\ref{PB}) around
the value of potential at the cell boundary,
%\cite{remark}}
$\Phi _{R}=\Phi(R_{WS})$, unknown at this point, which yields 
\begin{equation}
\nabla^2 \delta \Phi = \kappa_{\star}^2 \left(\gamma _{0}+\delta \Phi \right) 
\label{DH_WS}
\end{equation}
where $\delta \Phi \,=\,\Phi -\Phi _{R}$, 
$\kappa_{\star}^{2}=\kappa ^{2}{\rm cosh}(\Phi _{R})$ and 
$\gamma _{0}=\sqrt{1-(\kappa/\kappa_{\star})^{4}}$. Note that the relevant 
screening length $\kappa_{\star}^{-1}$ (always smaller than $\kappa ^{-1}$, 
a general feature for
finite concentration) is not a parameter, and should be determined at the
end of the calculation. Equation (\ref{DH_WS}) is supplemented by two
boundary conditions: the consistency constraint [$\delta \Phi (R_{WS})=0$]
and the global electroneutrality 
(which imposes a vanishing normal electric field at the WS boundary).
%($\nabla \Phi \left| _{R_{WS}}\right. =0$).
To generalize the approach discussed in the limit of infinite dilution,
we propose the following prescription (providing a third boundary 
condition): the {\it difference of potential} between the macro-ion and the 
WS surface is $\delta \Phi (a)=4$. Here again, the effective
charge is obtained from Gauss theorem at the macro-ion's surface.

This generalized procedure is now explicited for a solution of spherical
macro-ions with concentration $\rho $. The radius of the WS spherical cell is
given as $R_{WS}=(4\pi \rho /3)^{-1/3}$. In this geometry, the (LPB)
solution of Eq. (\ref{DH_WS}) reads 
\begin{equation}
\delta \Phi (r)=\gamma_0
\left[-1+f_{+}{\frac{e^{\kappa
_{\star}r}}{r}}+f_{-}{\frac{e^{-\kappa_{\star}r}}{r}}
\right]  \label{Sphere}
\end{equation}
where 
$f_{\pm }=(\kappa_{\star}a\pm 1)/(2\kappa_{\star})\exp (\mp \kappa_{\star}R_{WS})$.
Our prescription allows to compute $\kappa_{\star}$, such that $\delta \Phi (a)=4$. 
This equation is solved numerically for $%
\kappa_{\star}$
using a simple numerical Newton procedure. The effective charge follows from 
the gradient of 
$\delta \Phi(r)$ in Eq. (\ref{Sphere}) taken at $r=a$.
The corresponding $\Zsat$ as a function
of volume fraction $\eta =4\pi \rho a^{3}/3$ is displayed in Fig. 
\ref{Effective_Charge}, with a comparison to its counterpart deduced from the
numerical solution of PB theory supplemented with the popular procedure
proposed by Alexander et al \cite{Alex}. On this figure, we have also
plotted the results obtained without added salt [where the term ${\rm sinh}%
\Phi $ on the r.h.s. of Eq. (\ref{PB}) is replaced by $\exp \Phi$, due
to the absence of co-ions]. Our results are fully compatible with those
obtained from Alexander's method, with a similar agreement for cylindrical
macro-ions (not shown). It is eventually instructive to note that 
for a charged plane confined without added 
salt in a WS slab of width $2h$, Alexander's saturation surface charge may
be computed analytically with the result \cite{BTA}
$\sigma_{\hbox{\scriptsize sat}} = 2^{-3/2} \sinh(\pi /\sqrt{2}) /(\ell_{B} h)$
whereas we obtain 
$\sigma_{\hbox{\scriptsize sat}} = \sqrt{6} \hbox{Argcosh}(5)/(\pi \ell_{B} h)$
following our prescription. Both expressions agree within 10\% and remarkably 
exhibit the same functional dependence on $\ell_B$ and density (through $h$).

That our prescription compares favorably with Alexander's procedure 
for the planar, cylindrical and spherical geometries, 
calls for the more stringent test of
confronting our predictions against experimental data. We 
shall consider two specific
situations corresponding to two different geometries: crystallization of
charged colloidal suspensions and osmotic pressure in B-DNA solutions.
  
\begin{figure}[htb]
\psfig{file=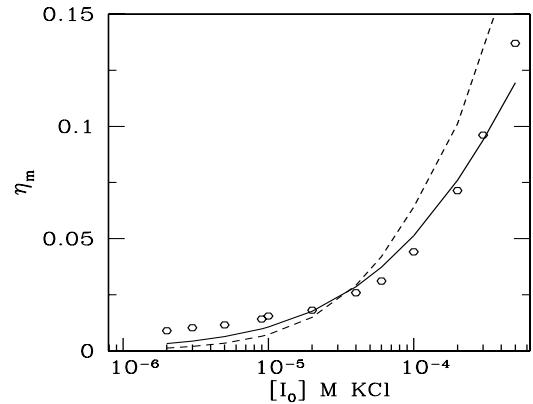,width=7.5cm,height=6.0cm}
\caption{ Liquid-solid transition of charged polystyrene colloids: 
volume fraction for melting $\eta_m$ 
as a function of salt ionic strength $I_0$. 
Dots are  experimental points for the melting line extracted from 
Ref. {\protect\cite{Gast}}. The solid line is the theoretical prediction 
for the melting transition using our prescription for effective charges.
The dashed line corresponds to $\Zeff=880$ (see text).}
\label{spheres}
\end{figure}

{\bf Crystallization of charged spheres}. Investigation into the phase diagram 
of charged polystyrene colloids has been conducted experimentally by Monovoukas and
Gast \cite{Gast} and  compared to the phase diagram of a
system where particles interact through a Yukawa potential (deduced from
extensive molecular dynamics simulations \cite{Robbins}). However this comparison 
requires an ad-hoc choice for 
$Z_{\hbox{\scriptsize eff}}$. We use here the  results 
we found for the effective charge as a function of ionic
strength, which we insert into the numerical generic phase diagram of Yukawa
systems \cite{Robbins}. We emphasize that
there is {\it no adjustable parameter} in our equations since the radius of the
polystyrene beads, the only parameter entering our description, was
independently measured to be $a=667\,$\AA . We only make the (reasonable)
assumption  that the bare charge $Z$ of the colloids is large enough 
to have $\Zeff \simeq \Zsat$.
The results for the melting line
using our prescription for the effective charge 
are confronted to the experimental data in Fig. 
\ref{spheres}. We also plot the result for the melting line for 
an ad-hoc constant
effective charge, $\Zeff=880$, as was proposed in \cite{Gast} 
(while in our case the latter varies 
between 500 and 2000 on the melting curve). The 
observed agreement of our 
results illustrates both the pertinence of our prescription for 
$Z_{\hbox{\scriptsize sat}}$ and the relevance of the PB saturation picture 
for macro-ions of large bare charge, for monovalent micro-ions in water.

%%%%%%%%Fig 3%%%%%%%%%%   
%%%%%%%%%%%%%%%%%%%%%%% 

{\bf Osmotic coefficient of B-DNA}. A similar test of our method may
be performed for the cylindrical geometry using the experimental results for
rigid cylindrical polyelectrolytes such as B-DNA \cite{BDNA}. We
specifically consider the measurements of the osmotic coefficient $\phi=\Pi_{%
\hbox{\scriptsize osm}}/\Pi_{c}$, defined as the ratio between the osmotic
pressure $\Pi_{\hbox{\scriptsize osm}}$ to the pressure $\Pi_c$ of
releasable counter-ions having bare density $c_c$ ($\Pi_{c}=k_BT c_{c}$).
%Such a quantity has been investigated recently on the basis of a cell model
%using the non-linear PB equation \cite{Hansen}. 
Within the WS model, B-DNA
macro-ions are confined into cylindrical cells, whose radius $R_{WS}$ is
related to the bare concentration of DNA counter-ions as $c_{c}=(\ell_{DNA}
\pi R_{WS}^2)^{-1}$, with $\ell_{DNA}=1.7\,$\AA\ the distance between
charges along DNA. The osmotic pressure is related to the densities of
micro-ions at the cell boundary: $\Pi_{\hbox{\scriptsize osm}%
}=k_BT(\rho_++\rho_--2 I_0)$ \cite{Hansen}, which can be recast in the form $%
\Pi_{\hbox{\scriptsize osm}}=k_B T (\kappa^2_{\star}-\kappa^2 )/(4 \pi \ell_B)$
introducing the screening factor $\kappa_{\star}$ defined previously. This
latter quantity is computed from our prescription, following the same lines
as for the spherical case [see Eq. (\ref{Sphere})]. 
%%%%%%%%Fig 4%%%%%%%%%%   
\begin{figure}[htb]
\psfig{file=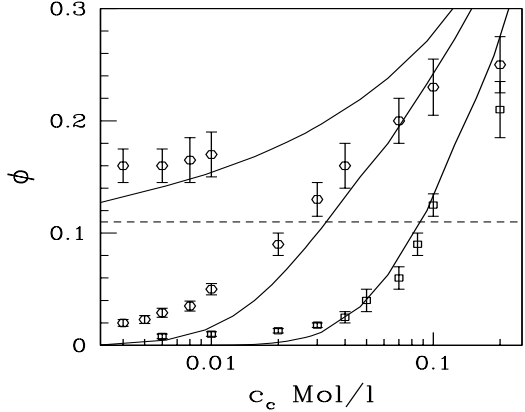,width=7.5cm,height=6.0cm}
\caption{Osmotic coefficient of B-DNA vs. density of DNA
phosphate ions $c_{c}$, for $I_0=$ 10 mM, 2mM and 0 mM (from
bottom to top). Dots: experiments of Refs. 
{\protect\cite{BDNA}}; solid lines: present prescription; 
dashed line: prediction of Oosawa-Manning condensation theory.}
\label{osm}
\end{figure}
%%%%%%%%%%%%%%%%%%%%%
In Fig. \ref{osm}, the
corresponding results for the osmotic coefficient are confronted against the
experimental data of Refs. \cite{BDNA}, showing again a good
quantitative agreement. As in Ref. \cite{Hansen}, we report the prediction
of classical Oosawa-Manning condensation theory, for which the
osmotic coefficient is constant [$\phi=\ell_{DNA}/(2 \ell_B)$]
at complete variance with the experiments. Again we emphasize that the only
quantity introduced in our description is the diameter ($a=10\,$\AA ) of DNA,
known from independent measurements.

In conclusion, we have put forward a simple method of asymptotic matching
to compute the effective charge at saturation for isolated macro-ions. In the 
situation of finite density, this method has been translated into a 
prescription, the validity of which has been assessed. 
This approach (mostly suited to describe the colloidal limit $\kappa a \gg 1$) 
amounts to consider the highly charged macro-ions as objects with
constant electrostatic potential $\sim 4kT/e$, independently of shape and
physico-chemical parameters (size, added 1:1 electrolyte\ldots ). 
As a general result we find that  the effective charge is  an 
{\it increasing} function of $\kappa $, which stems from the reduction of the
attraction between the counter-ions and the macro-ion. 
Addition of salt consequently brings two antagonist effects on the
effective Coulombic interaction between macro-ions: the range of the
interaction decreases due to screening, while the amplitude
increases due to the effective charge. The competition between these two
effects might be a key point in the understanding of these systems.

An important question concerning our approach, is that of the validity  
of PB theory for highly charged macro-ions. Within PB, 
micro-ions correlations are neglected, but the approach may 
still allow to
describe high macro/micro-ions couplings: in particular, for monovalent
micro-ions in water at room temperature, micro-ions correlations 
are negligible for all known macro-ions. This may 
no longer be the case in presence of
multi-valent micro-ions. More generally, 
PB is a reasonable approximation when the macro-ion
size $a$ is much larger than $\ell_B$ \cite{Groot}, and the saturation 
plateau of $\Zeff$ as a function of the bare charge 
$Z$ is an important physical phenomenon
that our approach allows to capture. When $a$ becomes of the same order
as $\ell_B$ the amount of counter-ion ``condensation'' 
found in molecular dynamics or
Monte Carlo simulations is larger than predicted by PB \cite{Groot,Deserno2}; 
our method for $\Zsat$ therefore provides an upper
bound for the effective charge. It is moreover noteworthy that for B-DNA
where $a/\ell_B \simeq 1.4$, our approach still gives a valuable first
approximation, and that omission of charge renormalization leads to spurious 
results (such as negative osmotic coefficients corresponding to an 
unphysical phase transition at physiological salt concentrations).

\end{multicols}


\begin{references}
\bibitem{Verwey}  E.J.W. Verwey and J.T.G. Overbeek, {\it Theory of the
Stability of Lyophobic Colloids} (Elsevier, Amsterdam, 1948).

\bibitem{Kjellander}  R. Kjellander in {\it Electrostatic effects in soft 
matter and biology}, C. Holm et al eds (Kluwer, Dordrecht, 2001). 

\bibitem{Manning} G.S. Manning, J. Chem. Phys. {\bf 51}, 924 (1969).

\bibitem{Belloni}  L. Belloni, Coll. and Surf. A  {\bf 140}, 227 (1998).

%\bibitem{Alex}  S. Alexander, P.M. Chaikin, P. Grant, G.J. Morales, and P.
%Pincus, J. Chem. Phys. {\bf 80}, 5776 (1984).
\bibitem{Alex}  S. Alexander {\it et al.}, J. Chem. Phys. {\bf 80}, 5776 (1984).

\bibitem{JPH}  J.-P. Hansen and H. L\"{o}wen, Annu. Rev. Phys. Chem. {\bf 51}%
, 209 (2000).

\bibitem{Diehl}  A. Diehl, M.C. Barbosa, and Y. Levin, Europhys. Lett. 
{\bf 53}, 80 (2001).

\bibitem{Gast}  Y. Monovoukas and A. P. Gast, J. Colloid and Interf. Sci. 
{\bf 128}, 533 (1989).

\bibitem{Robbins2}  M.J. Stevens, M.L. Falk, M.O. Robbins, J. Chem. Phys. 
{\bf 104}, 5209 (1996).

%\bibitem{Raman}  C.V. Ramanathan, J. Chem. Phys. {\bf 88} 3887 (1988).

\bibitem{Deserno}  M. Deserno, C. Holm, and S. May, Macromolecules {\bf 33},
199 (2000). 

\bibitem{Lukatsky} D.B. Lukatsky and S.A. Safran, 
Phys. Rev. E {\bf 63}, 011405. 


\bibitem{Rqe}  If on the other hand $Z$ is low, we have $\Zeff = Z$. 

\bibitem{BTA}  L. Bocquet, E. Trizac, and M. Aubouy, to appear in
J. Chem. Phys.

\bibitem{Crocker}  J.C. Crocker and D.G. Grier, Phys. Rev. Lett. {\bf 77},
1897 (1996).

\bibitem{Behrens}  S.H. Behrens and D.G. Grier, J. Chem. Phys. {\bf 115},
6716 (2001).

%\bibitem{Marcus}  R.A. Marcus, J. Chem. Phys. {\bf 23}, 1057 (1955).

%\bibitem{remark}  PB equation (\ref{PB}) in a cell describes a system in
%osmotic equilibrium with a salt reservoir of ionic strength $I_{0}=\kappa
%^{2}/(8\pi \ell _{B})$, with the further convention that $\Phi $ vanishes in
%the reservoir.


%\bibitem{Donnan}  M. Dubois, T. Zemb, L. Belloni, A. Delville, P. Levitz,
%and R. Setton, J. Chem. Phys. {\bf 75}, 944 (1992).
%\bibitem{Donnan}  M. Dubois {\it et al.}, J. Chem. Phys. {\bf 75}, 944 (1992).

\bibitem{Robbins}  M.O. Robbins, K. Kremer, G.S. Grest, J. Chem. Phys. {\bf %
88} 3286 (1988).

\bibitem{BDNA}  H.E. Auer and Z. Alexandrowicz, Biopolymers {\bf 8}, 1
(1969); E. Raspaud, L. da Conceicao, and F. Livolant, Phys. Rev. Lett. {\bf %
84}, 2533 (2000).

\bibitem{Hansen}  P.L. Hansen, R. Podgornik, and V.A. Parsegian, Phys. Rev.
E {\bf 64}, 021907 (2001).

\bibitem{Groot}
R.D. Groot, 
J. Chem. Phys. {\bf 95}, 9191 (1991). 

\bibitem{Deserno2}
M. Deserno and C. Holm, cond-mat/0203599.

\end{references}
\end{document}